\documentclass[aps,pra,twocolumn,showpacs,a4paper]{revtex4}
\usepackage{amsmath,amsfonts}
\usepackage{graphicx, subfigure}
\graphicspath{{figdistilled/}{./}}

\renewcommand{\vec}[1]{\ensuremath{\boldsymbol{#1}}}
\newcommand{\matr}[1]{\mathbf{{#1}}}

\newcommand{\id}{\ensuremath{\boldsymbol{1}}}
\newcommand{\ham}{\ensuremath{H}} 
\newcommand{\hilb}{\ensuremath{\mathcal{H}}}
\newcommand{\fid}{\ensuremath{\mathcal{F}}}
\newcommand{\fidt}{\ensuremath{\mathcal{T}}}
\newcommand{\be}{\ensuremath{\bra{\text{e}}}}
\newcommand{\bl}{\ensuremath{\bra{1}}}
\newcommand{\bo}{\ensuremath{\bra{0}}}
\newcommand{\ke}{\ensuremath{\ket{\text{e}}}}
\newcommand{\kl}{\ensuremath{\ket{1}}}
\newcommand{\ko}{\ensuremath{\ket{0}}}

\newcommand{\complex}{\ensuremath{\mathbb{C}}}

\newcommand{\dd}[1]{\frac{\partial}{\partial{#1}}}

\newcommand{\bra}[1]{\left\langle{#1}\right\rvert}
\newcommand{\ket}[1]{\left\lvert{#1}\right\rangle}

{\catcode`\|=\active
\gdef\Braket#1{\left<\mathcode`\|"8000\let|\BraVert {#1}\right>}}
\def\BraVert{\egroup\,\vrule\,\bgroup}
\newcommand{\braket}[1]{\Braket{#1}}

\DeclareMathOperator*{\mytrace}{Tr} 
\DeclareMathOperator*{\re}{Re} 
\DeclareMathOperator*{\im}{Im} 
\DeclareMathOperator{\sech}{sech}

\newcommand{\evol}{\mathbf{U}}

\newcommand{\ctrl}{\vec{\varepsilon}}
\newcommand{\cost}{\matr{\Lambda}}
\newcommand{\stat}{\matr{U}}
\newcommand{\statid}{\stat_0}
\newcommand{\stato}{\id}

\newcommand{\evalt}[1]{\left.{#1}\right\rvert_T}
\newcommand{\octH}{h}

\begin{document} 

\title{%
Designing robust gate implementations for quantum information processing
}
\date{\today} 
\author{Janus Wesenberg}
\affiliation{%
QUANTOP, Danish Research Foundation Center for Quantum Optics,
Institute of Physics and Astronomy, University of Aarhus, DK-8000
{\AA}rhus C, Denmark
}
\begin{abstract}
  Quantum information processing systems are often operated through
  time dependent controls; choosing these controls in a way that makes
  the resulting operation insensitive to variations in unknown or 
  uncontrollable system parameters is an important prerequisite for
  obtaining high-fidelity gate operations.
  In this article we present a numerical method for constructing such
  robust control sequences for a quite general class of quantum
  information processing systems.
  As an application of the method we have designed a robust
  implementation of a phase-shift operation central to rare earth
  quantum computing, an ensemble quantum computing system proposed by
  Ohlsson et.~al. \cite{ohlssona02:quant_comput_hardw_based_rare}.
  In this case the method has been used to obtain a
  high degree of insensitivity with respect to differences between
  ensemble members, but it is equally well suited for 
  quantum computing with a single physical system.
\end{abstract}


\maketitle

\section*{\label{sec:introduction}Introduction}
Many potential quantum information processing systems are controlled
by means of a set of time-dependent parameters, such as quasi-static
electromagnetic fields
\cite{nakamura99:coher_contr_macros_quant_states}, radio-frequency,
\cite{cory97:ensem_quant_comput_nmr_spect,kane98:silic_based_nuclear_spin_quant},
or optical fields
\cite{cirac95:quant_comput_with_cold_trapp,%
imamoglu99:quant_infor_proces_using_quant,%
lukin00:quant_entan_optic_contr_atom,ohlssona02:quant_comput_hardw_based_rare}.
For most such systems, it is relatively simple to device a set of
controls that implement a given evolution in an ideal situation.
Often, however, a more careful choice of controls can lead to an
implementation that is less sensitive to variations in unknown or
uncontrollable system parameters. Examples of such robust
implementations include system specific solutions such as the hot gate
for ion trap quantum computing which is insensitive to vibrational
excitations \cite{soerensen99:quant_comput_with_ions_therm}, as well
as more general techniques such as composite pulses, a technique
originating in NMR spectroscopy
\cite{cummins02:tackl_system_error_quant_logic}.

In this article we describe a numerical method for designing robust
controls for systems where the evolution is adequately described by a
possibly non-unitary evolution operator $\evol(t)$.
This form does not allow a general master equation formulation, but it
is sufficient to establish worst case behavior in many quantum
computing settings where the worst case effects of decoherence and
loss can be adequately modeled by a Schr{\"o}dinger equation with a
non-Hermitian Hamiltonian.

As an application of the method, we will consider the construction of
a robust phase shift operation for the rare earth quantum computing
(REQC) system
\cite{ohlssona02:quant_comput_hardw_based_rare,longdell02:exper_demon_quant_state_tomog},
which is based on rare-earth ions embedded in a cryogenic crystal.
In each ion, two metastable ground-state hyperfine levels, labeled $\ko$ and
$\kl$, serve as a qubit register which is manipulated via optical
transitions from both states to an inhomogeneously shifted excited
state $\ke$.
The REQC system is an ensemble quantum computing system and
macroscopic numbers of ions are manipulated in parallel, addressed by
the value of the inhomogeneous shift of their $\ke$-state.
To obtain a sufficient number of ions within each frequency channel,
it is necessary to operate on all ions within a finite range of
inhomogeneous shifts, and the main difficulty in operating the REQC
system is to achieve the same evolution for each of these ions
independent of their particular inhomogeneous shift.

This article is divided into two sections: in section
\ref{sec:design-robust-puls} we describe the method we have used to
design robust gate implementations; these results should be applicable
to a variety of quantum information processing systems.
In section \ref{sec:optimal-control-reqc} we present the results of
applying the method to a specific problem relating to the REQC system,
and show that it is indeed possible to obtain very high degrees of
robustness.

\section{Designing robust gate implementations}
\label{sec:design-robust-puls}

We consider a collection of quantum system which evolve according to a
set of time-dependent controls $\ctrl(t)$. In addition to $\ctrl$, the
single system Hamiltonian $\ham(\vec{\xi},\ctrl(t))$ depends on a
system specific set of uncontrollable or unknown parameters
$\vec{\xi}$, such as field strength or quantum numbers corresponding
to unused degrees of freedom.

The evolution of each system is governed by the Schr{\"o}dinger
equation,
\begin{equation}
  \label{eq:schroedsys}
  i \hbar \dd{t} \stat(t)= \ham(\vec{\xi},\ctrl(t))\, \stat(t), 
  \qquad  \stat(0)=\stato,
\end{equation}
where we will allow the Hamiltonian $\ham$ to include non-Hermitian
terms describing loss and decoherence.

Our goal is to choose a set of controls that lead to an evolution
$\evol(T)$ which is as close as possible to a given desired evolution
$\evol_0$ over a range of $\vec{\xi}$-values.
To quantify this, we introduce an objective functional
$J(\vec{\xi},\ctrl)$ which describes the performance of a set of
controls $\ctrl$ for a given value of $\vec{\xi}$.
By convention we take a low value of $J$ to indicate a good
performance, and the problem of finding a robust set of controls thus 
corresponds to minimizing
\begin{equation}
  \label{eq:JXdef}
  J_X(\ctrl)=\max_{\vec{\xi}\in X} J(\vec{\xi},\ctrl), 
\end{equation}
where $X$ is the set of $\vec{\xi}$-values for which we want the
implementation to perform well.
The conceptually simple approach we have taken to this problem is to
replace $X$ with a discrete subset $X' \subset X$, so that the
minimization of $J_{X'}$ has the form of a standard minimax problem,
which can be solved efficiently provided that we are able to calculate 
$\partial J_{X'}/\partial \ctrl$. Below we show how to achieve this by
methods from optimal control theory.

\subsection{Calculating $\partial J/\partial \ctrl$}
\label{sec:canon-optim-contr}
In this section we show how $\partial J/\partial \ctrl$ may be
calculated for a quite general class of objective functionals.  To
keep the notation simple and avoid unnecessary restrictions, we will
consider the following generalization of the Schr{\"o}dinger equation
\eqref{eq:schroedsys},
\begin{equation}
  \label{eq:dynsys}
  \dot{\stat}(t)=\vec{f}(\stat(t),\ctrl(t)),
  \qquad \stat(0)=\stato,
\end{equation}
determining the evolution of a complex-valued, time-dependent matrix
$\stat$ due to a set of real-valued, time-dependent controls,
$\ctrl(t)$.
We will consider objective functionals of the form 
\begin{equation}
  \label{eq:objective}
  J(\ctrl)=\phi(\stat(T))+\int_0^T l(\stat(t),\ctrl(t)) dt,
\end{equation}
where $\phi(\stat(T))$ is a real-valued function quantifying how close
the final state $\evol(T)$ is to our goal, and the real valued
function $l((\stat(t),\ctrl(t))$, referred to as a penalty function,
can be chosen to discourage the use of certain control values.
Our goal is to calculate $\partial J/\partial \ctrl$ subject to the
constraint that $\ctrl$ and $\stat$ obey Eq.  \eqref{eq:dynsys}. To
achieve this we introduce the modified objective functional:
\begin{equation}
  \label{eq:mod_objective_c}
  J'=J-\int_0^T \mytrace \left(
  \cost^\dagger \left[\dot{\stat}-\vec{f}(\stat,\ctrl)\right] 
  + \text{h.c.} \right) dt,
\end{equation}
where the complex time-dependent adjoint state matrix $\cost$ is in
effect a continuous set of Lagrange multipliers leaving $J'$
identical to $J$, provided that $\stat$ and $\ctrl$ obey Eq.
\eqref{eq:dynsys}.
If we require $\cost$ to obey the adjoint equations
\begin{equation}
  \label{eq:dotvld-frac-part}
  \dot{\cost}^\dag=-\frac{ \partial{\octH}}{\partial \stat} ,\qquad
  \cost^\dag(T)=\evalt{\frac{\partial \phi}{\partial \stat}},
\end{equation}
where $\stat$ and $\stat^\dagger$ should be considered as
independent with respect to the partial derivative
and $\octH$ is defined as
\begin{equation}
  \label{eq:hamiltondef_c}
  \octH(\ctrl,\stat,\cost)=l(\stat,\ctrl)
  +\mytrace \left(\cost^\dagger \vec{f}(\stat,\ctrl)+\text{h.c.}\right),
\end{equation}
we find by integration by parts that the differential of $J'$ is given
by \cite{luenberger79:introd_dynam_system}
\begin{equation}
  \label{eq:compldj}
  dJ'= \int_0^T
  \frac{\partial \octH}{\partial \ctrl} \; \delta\ctrl(t) \, dt,
\end{equation}
from which the derivatives of $J$ with respect to the parameters
used to parametrize $\ctrl$ can be calculated by the chain rule.

We now return to the case of a quantum system governed by the
Schr{\"o}dinger equation \eqref{eq:schroedsys}.
If we assume the penalty function $l$ to be independent of
$\stat$, the adjoint equations in this case are
\begin{equation}
  \label{eq:schroedadj}
    i \hbar \dd{t} \cost(t)= \ham^\dag(\vec{\xi},\ctrl(t))\, \cost(t) ,\qquad 
    \cost(T)=\evalt{\frac{\partial \phi}{\partial \stat^\dag}},
\end{equation}
and $dJ$ is given by \eqref{eq:compldj} with 
\begin{equation}
  \label{eq:complexgrad}
  \frac{\partial \octH}{\partial \ctrl} = 
  \frac{\partial l}{\partial \ctrl} + 
  \frac{2}{\hbar} \im \mytrace\left(
    \cost^\dagger \frac{\partial \ham}{\partial \ctrl} \stat
  \right).
\end{equation}
The role of the adjoint state and the adjoint equations is often
described as back-propagating the errors in achieving the desired
final state.
If $\ham$ is Hermitian, the boundary value for $\cost$ can be
optimized for numerical computation as shown in Appendix
\ref{sec:optimized-co-state}.

\subsection{Fidelity of quantum evolution}
\label{sec:fidel-quant-evol}

We will now discuss the choice of the function $\phi$, 
quantifying how well the obtained evolution $\evol(T)$ approximates
$\evol_0$. 
As we are concerned with quantum information processing we will assume
that all operations start out with an unknown state in the qubit
subspace $\hilb_Q$ of the full system Hilbert space $\hilb$, and that
this subspace is invariant under the ideal evolution $\evol_0$.
The function $\phi$ should not depend on the evolution of states
outside $\hilb_Q$, nor on collective phases on the states originating
in $\hilb_Q$.
A cautious choice of $\phi$ fulfilling these conditions could be based
on the worst case overlap fidelity
\cite{nielsen00:quant_comput_quant_infor}:
\begin{equation}
  \label{eq:fiddef}
  \fid=\min_{\ket{\psi}\in\hilb_Q} 
  \left|
    \braket{\psi| \evol_0^\dag \evol(T) | \psi}
  \right|,
\end{equation}
which measures the least possible overlap between the obtained output
state $\evol(T) \ket{\psi}$  and the ideal output $\evol_0 \ket{\psi}$
for initial states in $\hilb_Q$.
This fidelity measure has the desirable quality that both population
transfer from $\hilb_Q$ to $\hilb_Q^\perp$ and population transfer
completely out of $\hilb$, as described by a non-unitary
evolution, is counted as loss of fidelity.

From the point of view of optimal control, a significant drawback of
the worst case overlap fidelity, $\fid$, is that it is computationally
complicated \cite{wesenberg03:robus_quant_gates_a_archit}.
A computationally accessible fidelity measure which share many
appealing features with $\fid$ is the trace fidelity
\cite{palao02:quant_comput_optim_contr_algor},
\begin{equation}
  \label{eq:tracefiddef}
  \fidt=\frac{1}{n} \left|\mytrace_{\hilb_Q}\left(\evol_0^\dag\, \evol\right)\right|,
\end{equation}
where $n=\dim(\hilb_Q)$.
As shown in Appendix \ref{sec:trace-fidelity}, $\fidt$ is related to 
$\fid$ by the strict bound
\begin{equation}
  \label{eq:fidestimaterepeat}
  1-\fid \le n (1-\fidt),
\end{equation}
indicating that we can safely replace $\fid$ by $\fidt$ for numerical
computations on a few qubits at high fidelity.

For numerical calculations it is beneficial to use $\fidt^2$ rather
than $\fidt$ \cite{palao03:optim_contr_theor_unitar_trans}; in the
calculations presented in the next section we have used
$\phi=1-\fidt^2$, leading to an adjoint state boundary condition of 
\begin{equation}
  \label{eq:adjointbound}
  \cost(T)
  = \evalt{\frac{\partial \phi}{\partial \stat^\dag}} \,
  = -\frac{1}{n^2} \statid \;\mytrace(\statid^\dag \stat(T)),
\end{equation}
which can be directly computed.

\subsection{Minimization algorithms}
\label{sec:minim-proc}

One approach to minimizing $J$ is to directly solve the extremum
condition $\partial J/\partial \ctrl=0$ for $\ctrl$. This task is
significantly simplified if a penalty function proportional to
$\ctrl^2$ is introduced, $\octH=\lambda \ctrl^2+\octH_0$, so that the
extremum condition according to \eqref{eq:compldj} reads,
\begin{equation}
  \label{eq:directiterat}
  \ctrl(t) = -\frac{1}{2 \lambda}
  \frac{\partial \octH_0}{\partial \ctrl},
\end{equation}
which may be used as an iterative formula for calculating $\ctrl$.
Variations over this iterative approach give rise to the Krotov
\cite{tannor92} and Zhu-Rabitz
\cite{zhu98:unifor_rapid_conver_algor_quant,zhu98:rapid_monot_conver_iterat_algor}
algorithms which have been shown to have excellent convergence
properties, and have been successfully applied to optimal control of
unitary transformations for one set of parameters by the group of R.
Kosloff
\cite{palao03:optim_contr_theor_unitar_trans,palao02:quant_comput_optim_contr_algor}.
A unifying view of these direct methods can be found in Ref.
\cite{maday03:new_formul_monot_conver_quant}.

In the work presented here we have chosen to use an indirect
minimization algorithm: rather than trying to solve the extremum
condition directly, we have used the gradient information obtained
through Eq. \eqref{eq:compldj} as input for a general sequential
quadratic programming procedure based on a constrained quasi-Newton
method \cite{fletcher87:pract_method_optim,gill81:pract_optim,lin99:method_large_bound_const_optim}.
The primary advantage of this approach is that we have total freedom
to choose the parametrization of the controls, and can place arbitrary
bounds on these.  This allows us to more accurately model the fact
that the experimental limitations most often only distinguish
between possible and impossible controls: no possible controls are
significantly harder than others.
An explicit field strength limit also serves to introduce an
absolute scale on which to introduce decay strengths etc.

\section{Application to rare earth quantum computing}
\label{sec:optimal-control-reqc}
\newcommand{\fullr}[2]{\rgate_{{#1}\text{e}}({#2})} 
\newcommand{\osech}{\Omega_{\sech}}

\begin{figure*}[htb!]
  \subfigure[ $\sech$-pulses]{%
    \includegraphics[width=8.4cm]{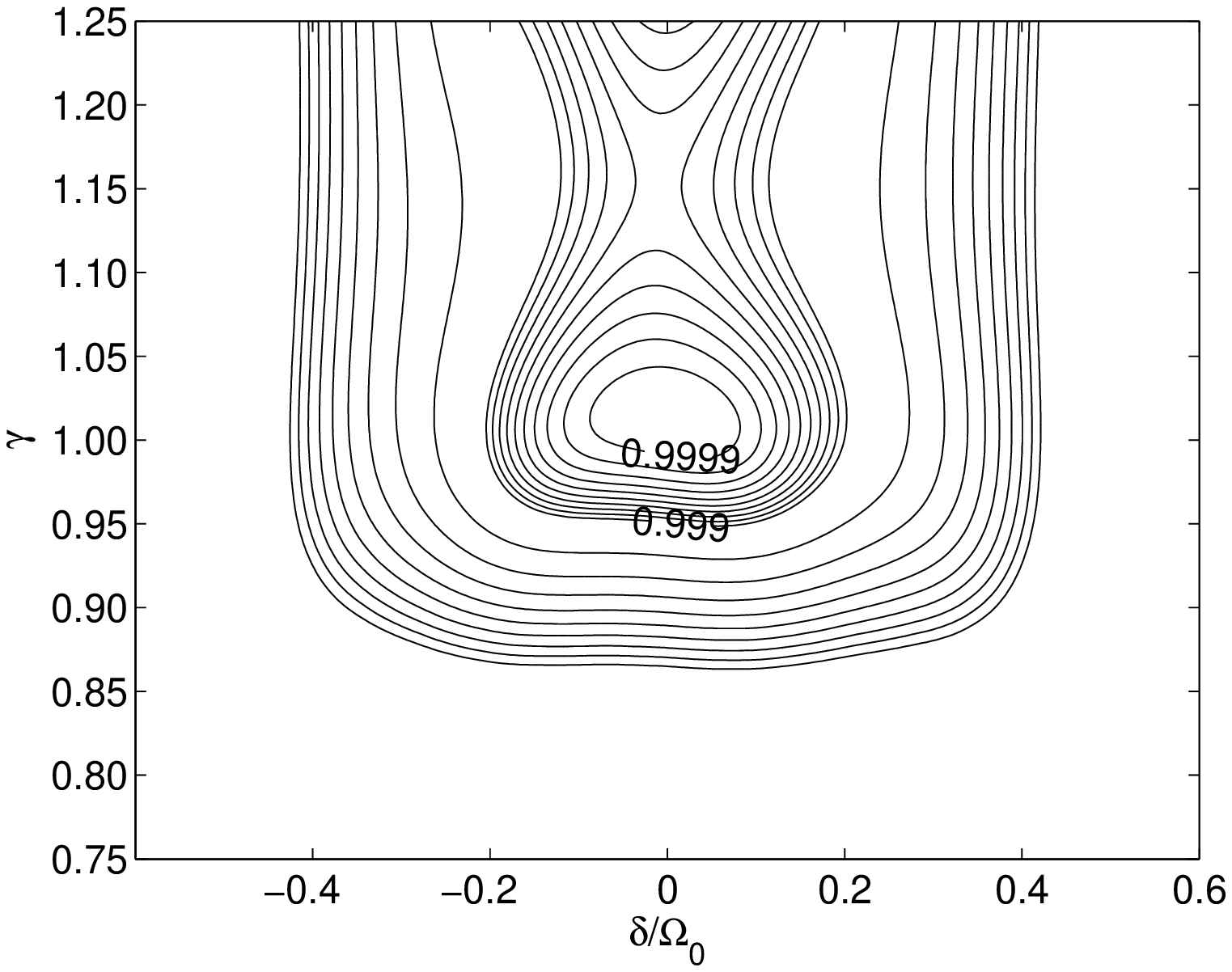}}
  \subfigure[ Optimized pulse]{%
    \includegraphics[width=8.4cm]{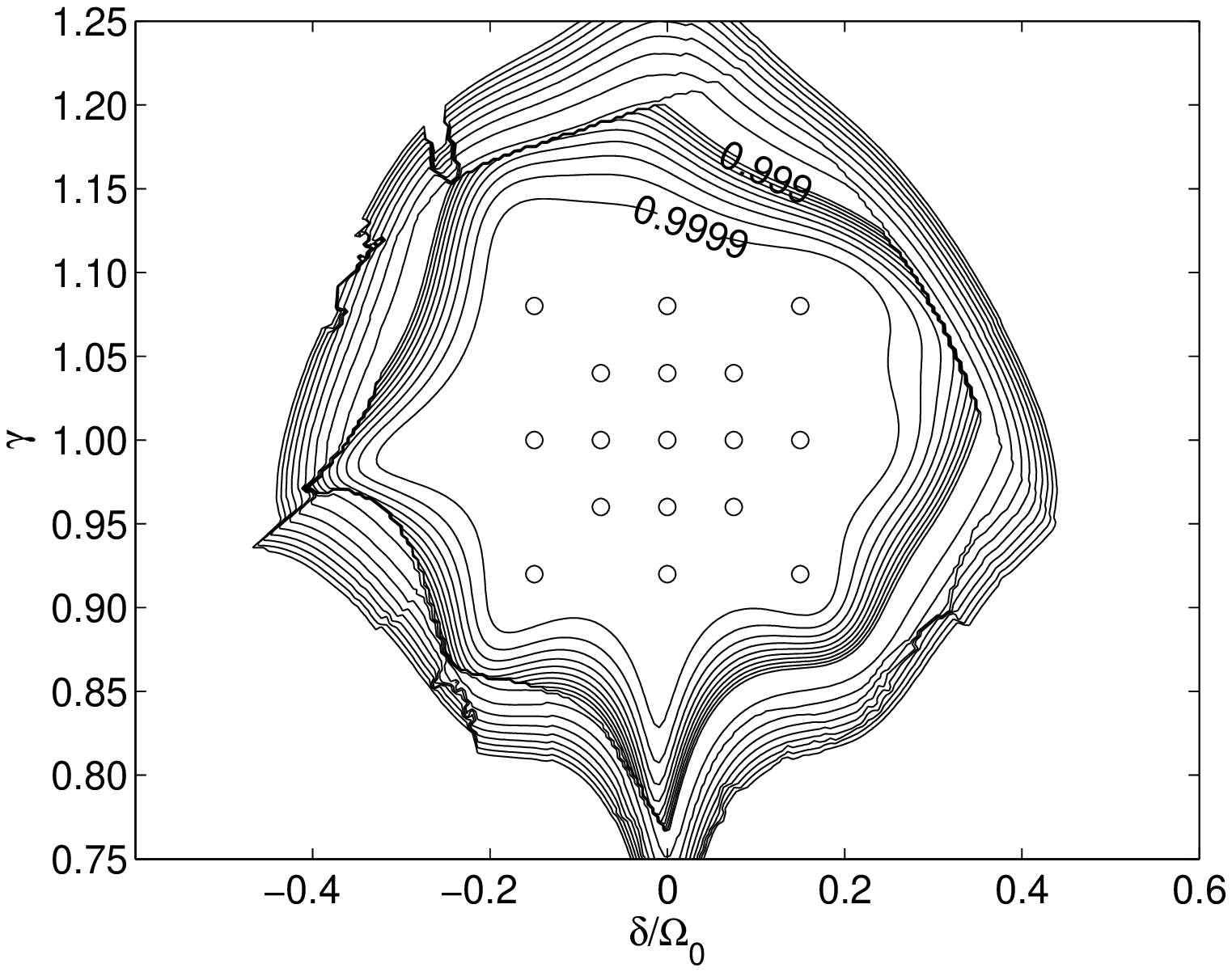}}
  \caption{%
    The worst case overlap fidelity of $\evol_0$ as given by Eq.
    \eqref{eq:phase0} implemented (a) by a series of $\sech$ pulses as
    suggested by Roos and M{\o}lmer
    \cite{roos03:robus_quant_comput_with_compos}, and (b) by an
    optimized pulse.
    The fidelity is plotted as a function of relative field strength
    $\gamma$ and inhomogeneous shift $\delta$ relative to the maximal
    resonant Rabi frequency $\Omega_0$. The duration of both pulses is
    $24 \pi/ \Omega_0$, and the circles in (b) indicates the parameter
    set $X'$ used during the optimization.  
  }
  \label{fig:fidelities}
\end{figure*}

\begin{figure}[htbp]
 \centering
 \includegraphics[width=8.4cm]{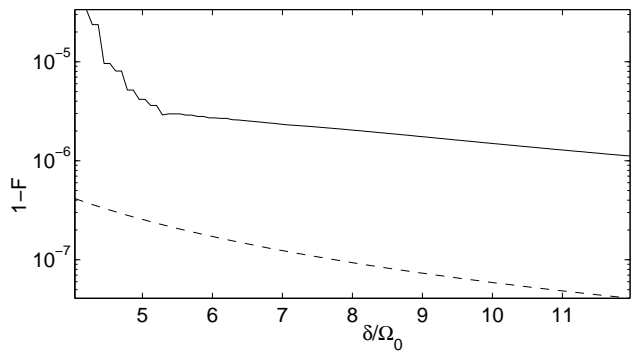}
 \caption{
   The effect of the $\sech$-pulse sequence (dashed line) and the
   optimized (solid line) implementation of $\evol_0$ on far-detuned
   ions, as described by $1-\fid$, where $\fid$ is calculated with
   respect to the identity, as these ions should ideally not be
   disturbed.
   Both implementations achieve worst case fidelities very close to
   unity, with the $\sech$-pulse sequence achieving the best results.
   The curve for the optimized pulse is actually a running maximum, as
   the actual value of $1-\fid$ oscillates with a period of $2 \pi/T$.
 }
  \label{fig:3levavoid_edge}
\end{figure}

The motivation for the work presented in this article has been the
design of robust gate implementations for the REQC system mentioned in
the introduction.
As an example, we will consider the construction of a robust
implementation of the single qubit operation
\begin{equation}
  \label{eq:phase0}
  \evol_0=\kl\bl - \ko\bo,
\end{equation}
which could simply be implemented by a single $2 \pi$ pulse on the
$\ko$-$\ke$ transition if we were not concerned with robustness.
Our primary concern will be to make the implementation robust with
respect to variations in the inhomogeneous shift $\delta$ of the
$\ke$-state in order to allow the use of finite with channels.
Since it is experimentally difficult to obtain a homogeneous field
strength over the crystal, we would also prefer the implementation
to be insensitive to variations in the relative field strength, which
we will denote $\gamma$.

In addition to requiring the implementation of $\evol_0$ to be robust
with respect to variations in $\delta$ and $\gamma$, we will add the
requirement that ions outside the channels should not be affected, as
this allows us to use the obtained implementation of $\evol_0$ as a
part of a controlled phase shift operation
\cite{lukin00:quant_entan_optic_contr_atom,
  ohlssona02:quant_comput_hardw_based_rare}: If the $\ke$-states of
the qubit ion and a controlling ion are coupled sufficiently strongly
by static dipole interaction, an excitation of the controlling ion
will effectively shift the qubit ion out of the channel thus
conditioning the evolution of the qubit on the state of the
controlling ion.

Even though the simplest implementation of $\evol_0$ for an ideal ion
with $\delta=0$ and $\gamma=1$ would involve only the $\ko$ and $\ke$
levels, a robust implementation must also involve the $\kl$-state since
the coupling of $\ko$ to $\ke$ will result in a $\delta$-dependent
phase on $\ko$ which can only be compensated by introducing the same
phase on the $\kl$-state, e.g. through phase compensating rotations
\cite{wesenberg03:robus_quant_gates_a_archit}.
A highly successful example of this approach is the $\sech$-pulse
sequence suggested by Roos and M{\o}lmer  
\cite{roos03:robus_quant_comput_with_compos},
which as illustrated by Fig. \ref{fig:fidelities}(a) yields a very
robust implementation, achieving high fidelities over a wide range of
parameter values.

We model the REQC system by the single ion Hamiltonian
\begin{equation}
  \label{eq:reqchamilton}
  \ham=-\delta \ke\be + \frac{\gamma}{2} \sum_{i=0,1} 
  \left(\Omega_i(t) \ke\bra{i} +\text{h.c.}\right),
\end{equation}
which does not include any effects of decay or decoherence.
In the notation introduced in section \ref{sec:design-robust-puls},
the system parameters are $\vec{\xi}=(\gamma,\delta)$, the controls
are $\ctrl=(\Omega_0,\Omega_1)$, and the qubit subspace is
$\hilb_Q=\{\ko,\kl\}$.
No penalty function is used: we use $J=\phi=1-\fidt^2$, and limit the field
by strict bounds on $\Omega_i(t)$, as this is the relevant limiting
parameter in the REQC system.
Inspired by the success of the $\sech$-based solution and the hat-like
Fourier spectrum of the $\sech$-pulse, $\Omega_i(t)$ is parametrized
in terms of a truncated Fourier basis.
Based on trial and error we have arrived at $49$ $\vec{\xi}$-values
to constitute $X'$, some within the neighborhood of $\delta=0$,
$\gamma=1$, and some at large detunings where the ions should not be
disturbed.
The result of the optimization with this choice of $X'$ is shown in Fig.
\ref{fig:fidelities}(b), where the circles indicate the members of $X'$.
It is evident from the plot that the optimization has achieved a high
fidelity over an even larger range of parameters than the
$\sech$-pulse sequence illustrated in Fig. \ref{fig:fidelities}(a).

With respect to not disturbing the detuned ions, both the optimized
pulse and the $\sech$-pulses obtain fidelities within $10^{-5}$ of
unity for $|\delta|>5 \Omega_0$, where $\Omega_0$ is the maximal
resonant Rabi frequency at $\gamma=1$.  As illustrated by Fig.
\ref{fig:3levavoid_edge}, which only shows fidelities at $\gamma=1$ as
$\evol(T)$ is nearly independent of $\gamma$ at $|\delta|\gg\Omega_0$,
the $\sech$-pulse sequence performs better than the optimized pulse in
this regime.

\section{Conclusions and outlook}
\label{sec:conclusions-outlook}
We have shown that it is possible to construct highly robust gate
implementations for quantum information processing by a quite general
method.
In particular, the method has been used to greatly enhance the
performance of a gate implementation for a model REQC system
by extending the range of inhomogeneous shifts and relative field
strengths over which an acceptable performance is achieved.

The model REQC system used in this article ignores many
performance degrading factors, the two most important being
decoherence and implementation noise.
Decoherence could in the present case be adequately modeled by a
non-Hermitian Hamiltonian, for which we expect
the method described in this paper to be able to find a robust
implementation as in the decoherence-free case.
It is not clear, however, how the method could be extended to address
the problem of robustness with respect to implementation imperfections.

\begin{acknowledgments}
  The author would like to thank the people at the Centre for Quantum
  Computer Technology at the University of Queensland for their
  hospitality, and Klaus M{\o}lmer for valuable comments on the
  manuscript.  This research was funded by project ESQUIRE of the
  IST-FET programme of the EC.
\end{acknowledgments}

\appendix

\section{The trace fidelity}
\label{sec:trace-fidelity}

In this section we prove the relation \eqref{eq:fidestimaterepeat}
between the worst case overlap fidelity, $\fid$, and the trace
fidelity $\fidt$.
Referring to the definition \eqref{eq:fiddef}, we note that $\fid$ is
completely determined by the restriction $\matr{O}$ of the operator 
$\evol_0^\dag \evol(T)$ to $\hilb_Q$. 
Since $\evol(T)$ describes the evolution of a quantum system, it is
possible to extend it to a unitary operation on a Hilbert space
containing $\hilb$, and $\matr{O}$ is consequently the
restriction of a unitary operator to $\hilb_Q$.
In the ideal case $\matr{O}$ will be equal to the identity on
$\hilb_Q$, perhaps with the exception of a complex phase.

$\fid$ is defined as the minimum of the overlap
$|\braket{\psi|\matr{O}|\psi}|$. Since the unit sphere of $\complex^n$
is compact, this minimum will be attained for some $\ket{\psi_0}$:
$\fid=|\braket{\psi_0|\matr{O}|\psi_0}|$.
We now extend $\{\ket{\psi_0}\}$ to an orthonormal basis
$\{\ket{\psi_k}\}_{k=0,\ldots,n-1}$ by the Gram-Schmidt
process. Evaluating the trace fidelity in this basis we find by the
triangle inequality:
\begin{subequations}
  \begin{align}
    \label{eq:proof}
    \fidt&\le\frac{1}{n} 
      \sum_{k=0}^{n-1} \left| \braket{\psi_k|\matr{O}|\psi_k}
    \right|\\
    &\le \frac{1}{n}\left(\fid +(n-1) \right),
    \label{eq:le-frac1nleftfid-+n}
  \end{align}
\end{subequations}
where we have used that $|\braket{\psi|\matr{O}|\psi}|\le 1$ for all
$\ket{\psi}$ since $\matr{O}$ is the restriction of a unitary
operator. By rewriting \eqref{eq:le-frac1nleftfid-+n} we obtain the
desired relation, $1-\fid \le n (1-\fidt)$.

We note that the established bound is strict in the sense that for any
$0 \le \fid_0 \le 1$, the operator
\begin{equation}
  \label{eq:boundop}
  \matr{O}_\fid = \id - (1-\fid_0) \ket{\psi}\bra{\psi}
\end{equation}
will fulfill Eq. \eqref{eq:fidestimaterepeat} with equality for any $\ket{\psi}$.

\section{Optimized adjoint state boundary condition}
\label{sec:optimized-co-state}
\newcommand{\statc}{\vec{x}}
\newcommand{\costc}{\vec{\lambda}}
In the case of a Hermitian Hamiltonian and a penalty function $l$,
that does not depend on the state $\stat$, it is possible to modify
the adjoint state boundary condition \eqref{eq:adjointbound} to reduce the
required accuracy of the adjoint state propagation.

In this case, we find according to Eqs. \eqref{eq:compldj} and \eqref{eq:complexgrad}
that $dJ$ has the form 
\begin{equation}
  \label{eq:dJdu}
  dJ=\int_0^T \left[
  \frac{\partial l}{\partial \ctrl}+
  2 \im\left( \sum_k
    \costc_k^\dag \frac{\partial \ham}{\partial \ctrl} \statc_k
  \right)
  \right] \delta \ctrl(t) \, dt, 
\end{equation}
where $\statc_k$ and $\costc_k$ denotes the $k$-th columns of $\stat$
and $\cost$ respectively.
Since $\ham$ and thus $\partial \ham/ \partial \ctrl$ are assumed to be
Hermitian, $dJ$ as given by \eqref{eq:dJdu} is not
affected by adding to $\costc_k$ a component of $\alpha_k \statc_k$ for
any real $\alpha_k$.
Furthermore, since $\costc_k$ and $\statc_k$ evolve according to the
same Schr\"odinger equation, this corresponds to modifying the
boundary condition for the adjoint state to read:
\begin{equation}
  \label{eq:adjointmodbound}
  \costc_k(T)
  = \frac{\partial\phi}{\partial \statc_k^\dag}+ 
  \alpha_k \statc_k(T), 
\end{equation}
for any real $\alpha_k$.

The obvious use of the freedom in the choice of boundary value is to
minimize the norm of the adjoint state, in order to relax the
requirements of the relative precision of the adjoint state
propagation.
This minimum is easily calculated from \eqref{eq:adjointmodbound}, but
we prefer to illustrate the physical background of the result by
calculating it in a different way:
The freedom in the choice of boundary value \eqref{eq:adjointmodbound} is
allowed by the Hermiticity of $\ham$. The same Hermiticity ensures
that $\statc_k$ is normalized, so that $\phi\circ\vec{n}(\statc_k)$,
where $\vec{n}$ is the normalization function $\vec{n}(\vec{v}) =
\vec{v}/|\vec{v}|$, is equal to $\phi(\statc_k)$.
The gradient of $\tilde{\phi}=\phi\circ\vec{n}$, however, is different
from that of $\phi$. In fact we find
\begin{equation}
  \label{eq:modboundgeneral}
  \frac{\partial\tilde{\phi}}{\partial \statc_k^\dag} =
  \frac{\partial\phi}{\partial \statc_k^\dag} -
  \re\left[ 
    \statc_k^\dag \frac{\partial\phi}{\partial \statc_k^\dag}
  \right]\; \statc_k,
\end{equation}
where $\statc_k$ and $\statc_k^\dag$ should be considered independent
with respect to the derivative.
Comparing this expression to Eq. \eqref{eq:adjointmodbound}, it is
tempting to let 
$\evalt{\alpha_k=\re(\statc_k^\dag \partial\phi/\partial \statc_k^\dag)}$, 
which is indeed the answer found by  minimizing $|\statc_k(T)|$ subject to
Eq. \eqref{eq:adjointmodbound}.
 
The modified boundary condition of Eq. \eqref{eq:modboundgeneral}
carries the same error information as that of Eq.
\eqref{eq:adjointbound}, but the required relative numerical precision
when propagating the adjoint state will be significantly reduced.


\begin{thebibliography}{23}
\expandafter\ifx\csname natexlab\endcsname\relax\def\natexlab#1{#1}\fi
\expandafter\ifx\csname bibnamefont\endcsname\relax
  \def\bibnamefont#1{#1}\fi
\expandafter\ifx\csname bibfnamefont\endcsname\relax
  \def\bibfnamefont#1{#1}\fi
\expandafter\ifx\csname citenamefont\endcsname\relax
  \def\citenamefont#1{#1}\fi
\expandafter\ifx\csname url\endcsname\relax
  \def\url#1{\texttt{#1}}\fi
\expandafter\ifx\csname urlprefix\endcsname\relax\def\urlprefix{URL }\fi
\providecommand{\bibinfo}[2]{#2}
\providecommand{\eprint}[2][]{\url{#2}}

\bibitem[{\citenamefont{Ohlsson et~al.}(2002)\citenamefont{Ohlsson, Mohan, and
  Kr{\"o}ll}}]{ohlssona02:quant_comput_hardw_based_rare}
\bibinfo{author}{\bibfnamefont{N.}~\bibnamefont{Ohlsson}},
  \bibinfo{author}{\bibfnamefont{R.~K.} \bibnamefont{Mohan}}, \bibnamefont{and}
  \bibinfo{author}{\bibfnamefont{S.}~\bibnamefont{Kr{\"o}ll}},
  \bibinfo{journal}{Opt. Comm.} \textbf{\bibinfo{volume}{201}},
  \bibinfo{pages}{71} (\bibinfo{year}{2002}).

\bibitem[{\citenamefont{Nakamura et~al.}(1999)\citenamefont{Nakamura, Pashkin,
  and Tsai}}]{nakamura99:coher_contr_macros_quant_states}
\bibinfo{author}{\bibfnamefont{Y.}~\bibnamefont{Nakamura}},
  \bibinfo{author}{\bibfnamefont{Y.~A.} \bibnamefont{Pashkin}},
  \bibnamefont{and} \bibinfo{author}{\bibfnamefont{J.~S.} \bibnamefont{Tsai}},
  \bibinfo{journal}{Nature} \textbf{\bibinfo{volume}{398}},
  \bibinfo{pages}{786} (\bibinfo{year}{1999}).

\bibitem[{\citenamefont{Cory et~al.}(1997)\citenamefont{Cory, Fahmy, and
  Havel}}]{cory97:ensem_quant_comput_nmr_spect}
\bibinfo{author}{\bibfnamefont{D.~G.} \bibnamefont{Cory}},
  \bibinfo{author}{\bibfnamefont{A.~F.} \bibnamefont{Fahmy}}, \bibnamefont{and}
  \bibinfo{author}{\bibfnamefont{T.~F.} \bibnamefont{Havel}},
  \bibinfo{journal}{Proc. Natl. Acad. Sci.} \textbf{\bibinfo{volume}{94}},
  \bibinfo{pages}{1634} (\bibinfo{year}{1997}).

\bibitem[{\citenamefont{Kane}(1998)}]{kane98:silic_based_nuclear_spin_quant}
\bibinfo{author}{\bibfnamefont{B.~E.} \bibnamefont{Kane}},
  \bibinfo{journal}{Nature} \textbf{\bibinfo{volume}{393}},
  \bibinfo{pages}{133} (\bibinfo{year}{1998}).

\bibitem[{\citenamefont{Cirac and
  Zoller}(1995)}]{cirac95:quant_comput_with_cold_trapp}
\bibinfo{author}{\bibfnamefont{J.~I.} \bibnamefont{Cirac}} \bibnamefont{and}
  \bibinfo{author}{\bibfnamefont{P.}~\bibnamefont{Zoller}},
  \bibinfo{journal}{Phys. Rev. Lett.} \textbf{\bibinfo{volume}{74}},
  \bibinfo{pages}{4091} (\bibinfo{year}{1995}).

\bibitem[{\citenamefont{Imamoglu et~al.}(1999)\citenamefont{Imamoglu,
  Awschalom, Burkard, Di-Vincenzo, Loss, Sherwin, and
  Small}}]{imamoglu99:quant_infor_proces_using_quant}
\bibinfo{author}{\bibfnamefont{A.}~\bibnamefont{Imamoglu}},
  \bibinfo{author}{\bibfnamefont{D.~D.} \bibnamefont{Awschalom}},
  \bibinfo{author}{\bibfnamefont{G.}~\bibnamefont{Burkard}},
  \bibinfo{author}{\bibfnamefont{D.~P.} \bibnamefont{Di-Vincenzo}},
  \bibinfo{author}{\bibfnamefont{D.}~\bibnamefont{Loss}},
  \bibinfo{author}{\bibfnamefont{M.}~\bibnamefont{Sherwin}}, \bibnamefont{and}
  \bibinfo{author}{\bibfnamefont{A.}~\bibnamefont{Small}},
  \bibinfo{journal}{Phys. Rev. Lett.} \textbf{\bibinfo{volume}{83}},
  \bibinfo{pages}{4204} (\bibinfo{year}{1999}).

\bibitem[{\citenamefont{Lukin and
  Hemmer}(2000)}]{lukin00:quant_entan_optic_contr_atom}
\bibinfo{author}{\bibfnamefont{M.~D.} \bibnamefont{Lukin}} \bibnamefont{and}
  \bibinfo{author}{\bibfnamefont{P.~R.} \bibnamefont{Hemmer}},
  \bibinfo{journal}{Phys. Rev. Lett.} \textbf{\bibinfo{volume}{84}},
  \bibinfo{pages}{2818} (\bibinfo{year}{2000}).

\bibitem[{\citenamefont{S{\o}rensen and
  M{\o}lmer}(1999)}]{soerensen99:quant_comput_with_ions_therm}
\bibinfo{author}{\bibfnamefont{A.}~\bibnamefont{S{\o}rensen}} \bibnamefont{and}
  \bibinfo{author}{\bibfnamefont{K.}~\bibnamefont{M{\o}lmer}},
  \bibinfo{journal}{Phys. Rev. Lett.} \textbf{\bibinfo{volume}{82}},
  \bibinfo{pages}{1971} (\bibinfo{year}{1999}).

\bibitem[{\citenamefont{Cummins et~al.}(2003)\citenamefont{Cummins, Llewellyn,
  and Jones}}]{cummins02:tackl_system_error_quant_logic}
\bibinfo{author}{\bibfnamefont{H.~K.} \bibnamefont{Cummins}},
  \bibinfo{author}{\bibfnamefont{G.}~\bibnamefont{Llewellyn}},
  \bibnamefont{and} \bibinfo{author}{\bibfnamefont{J.~A.} \bibnamefont{Jones}},
  \bibinfo{journal}{Phys. Rev. A} \textbf{\bibinfo{volume}{67}}
  (\bibinfo{year}{2003}).

\bibitem[{\citenamefont{Longdell and
  Sellars}(2002)}]{longdell02:exper_demon_quant_state_tomog}
\bibinfo{author}{\bibfnamefont{J.~J.} \bibnamefont{Longdell}} \bibnamefont{and}
  \bibinfo{author}{\bibfnamefont{M.~J.} \bibnamefont{Sellars}}
  (\bibinfo{year}{2002}), \eprint{quant-ph/0208182}.

\bibitem[{\citenamefont{Luenberger}(1979)}]{luenberger79:introd_dynam_system}
\bibinfo{author}{\bibfnamefont{D.~G.} \bibnamefont{Luenberger}},
  \emph{\bibinfo{title}{Introduction to Dynamic Systems}}
  (\bibinfo{publisher}{John Wiley \& Sons}, \bibinfo{year}{1979}).

\bibitem[{\citenamefont{Nielsen and
  Chuang}(2000)}]{nielsen00:quant_comput_quant_infor}
\bibinfo{author}{\bibfnamefont{M.~A.} \bibnamefont{Nielsen}} \bibnamefont{and}
  \bibinfo{author}{\bibfnamefont{I.~L.} \bibnamefont{Chuang}},
  \emph{\bibinfo{title}{Quantum Computation and Quantum Information}}
  (\bibinfo{publisher}{Cambridge University Press}, \bibinfo{year}{2000}).

\bibitem[{\citenamefont{Wesenberg and
  M{\o}lmer}(2003)}]{wesenberg03:robus_quant_gates_a_archit}
\bibinfo{author}{\bibfnamefont{J.}~\bibnamefont{Wesenberg}} \bibnamefont{and}
  \bibinfo{author}{\bibfnamefont{K.}~\bibnamefont{M{\o}lmer}},
  \bibinfo{journal}{Phys. Rev. A} \textbf{\bibinfo{volume}{68}},
  \bibinfo{pages}{012320} (\bibinfo{year}{2003}).

\bibitem[{\citenamefont{Palao and
  Kosloff}(2002)}]{palao02:quant_comput_optim_contr_algor}
\bibinfo{author}{\bibfnamefont{J.~P.} \bibnamefont{Palao}} \bibnamefont{and}
  \bibinfo{author}{\bibfnamefont{R.}~\bibnamefont{Kosloff}},
  \bibinfo{journal}{Phys. Rev. Lett.} \textbf{\bibinfo{volume}{89}},
  \bibinfo{pages}{188301} (\bibinfo{year}{2002}).

\bibitem[{\citenamefont{Palao and
  Kosloff}(2003)}]{palao03:optim_contr_theor_unitar_trans}
\bibinfo{author}{\bibfnamefont{J.~P.} \bibnamefont{Palao}} \bibnamefont{and}
  \bibinfo{author}{\bibfnamefont{R.}~\bibnamefont{Kosloff}}
  (\bibinfo{year}{2003}), \eprint{quant-ph/0309011}.

\bibitem[{\citenamefont{Tannor et~al.}(1992)\citenamefont{Tannor, Kazakov, and
  Orlov}}]{tannor92}
\bibinfo{author}{\bibfnamefont{D.~J.} \bibnamefont{Tannor}},
  \bibinfo{author}{\bibfnamefont{V.~A.} \bibnamefont{Kazakov}},
  \bibnamefont{and} \bibinfo{author}{\bibfnamefont{V.}~\bibnamefont{Orlov}}, in
  \emph{\bibinfo{booktitle}{Time-Dependent Quantum Molecular Dynamics}}, edited
  by \bibinfo{editor}{\bibfnamefont{J.}~\bibnamefont{Broeckhove}}
  \bibnamefont{and}
  \bibinfo{editor}{\bibfnamefont{L.}~\bibnamefont{Lathouwers}}
  (\bibinfo{publisher}{Plenum, New York}, \bibinfo{year}{1992}), pp.
  \bibinfo{pages}{347--360}.

\bibitem[{\citenamefont{Zhu and
  Rabitz}(1998{\natexlab{a}})}]{zhu98:unifor_rapid_conver_algor_quant}
\bibinfo{author}{\bibfnamefont{W.}~\bibnamefont{Zhu}} \bibnamefont{and}
  \bibinfo{author}{\bibfnamefont{H.}~\bibnamefont{Rabitz}},
  \bibinfo{journal}{Phys. Rev. A} \textbf{\bibinfo{volume}{58}},
  \bibinfo{pages}{4741} (\bibinfo{year}{1998}{\natexlab{a}}).

\bibitem[{\citenamefont{Zhu and
  Rabitz}(1998{\natexlab{b}})}]{zhu98:rapid_monot_conver_iterat_algor}
\bibinfo{author}{\bibfnamefont{W.}~\bibnamefont{Zhu}} \bibnamefont{and}
  \bibinfo{author}{\bibfnamefont{H.}~\bibnamefont{Rabitz}},
  \bibinfo{journal}{Jour. Chem. Phys.} \textbf{\bibinfo{volume}{109}},
  \bibinfo{pages}{385} (\bibinfo{year}{1998}{\natexlab{b}}).

\bibitem[{\citenamefont{Maday and
  Turinici}(2003)}]{maday03:new_formul_monot_conver_quant}
\bibinfo{author}{\bibfnamefont{Y.}~\bibnamefont{Maday}} \bibnamefont{and}
  \bibinfo{author}{\bibfnamefont{G.}~\bibnamefont{Turinici}},
  \bibinfo{journal}{J. Chem. Phys.} \textbf{\bibinfo{volume}{118}},
  \bibinfo{pages}{8191} (\bibinfo{year}{2003}).

\bibitem[{\citenamefont{Fletcher}(1987)}]{fletcher87:pract_method_optim}
\bibinfo{author}{\bibfnamefont{R.}~\bibnamefont{Fletcher}},
  \emph{\bibinfo{title}{Practical Methods of Optimization}}
  (\bibinfo{publisher}{John Wiley {\&} Sons}, \bibinfo{year}{1987}),
  \bibinfo{edition}{2nd} ed.

\bibitem[{\citenamefont{Gill et~al.}(1981)\citenamefont{Gill, Murray, and
  Wright}}]{gill81:pract_optim}
\bibinfo{author}{\bibfnamefont{P.~E.} \bibnamefont{Gill}},
  \bibinfo{author}{\bibfnamefont{W.}~\bibnamefont{Murray}}, \bibnamefont{and}
  \bibinfo{author}{\bibfnamefont{M.~H.} \bibnamefont{Wright}},
  \emph{\bibinfo{title}{Practical Optimization}} (\bibinfo{publisher}{Academic
  Press}, \bibinfo{year}{1981}).

\bibitem[{\citenamefont{Lin and
  Mor{\'e}}(1999)}]{lin99:method_large_bound_const_optim}
\bibinfo{author}{\bibfnamefont{C.-J.} \bibnamefont{Lin}} \bibnamefont{and}
  \bibinfo{author}{\bibfnamefont{J.~J.} \bibnamefont{Mor{\'e}}},
  \bibinfo{journal}{SIAM Journal on Optimization} \textbf{\bibinfo{volume}{9}},
  \bibinfo{pages}{1100} (\bibinfo{year}{1999}).

\bibitem[{\citenamefont{Roos and
  M{\o}lmer}(2003)}]{roos03:robus_quant_comput_with_compos}
\bibinfo{author}{\bibfnamefont{I.}~\bibnamefont{Roos}} \bibnamefont{and}
  \bibinfo{author}{\bibfnamefont{K.}~\bibnamefont{M{\o}lmer}}
  (\bibinfo{year}{2003}), \eprint{quant-ph/0305060}.

\end{thebibliography}

\end{document}